\documentclass[12pt]{article}
\usepackage[pctex32]{graphics}
\begin{document}
\vspace{2cm}
\begin{center}
~
\\
~
{\bf  \Large Holographic Gauge Theory with Maxwell Magnetic Field}
\vspace{1cm}

                      Wung-Hong Huang\\
                       Department of Physics\\
                       National Cheng Kung University\\
                       Tainan,Taiwan\\

\end{center}
\vspace{2cm}
We first apply the transformation of mixing azimuthal with wrapped coordinate to the 11D M-theory with a stack N M5-branes to find the spacetime of a stack of  N D4-branes with magnetic field in 10D IIA string theory, after the Kaluza-Klein reduction.  In the near-horizon limit the background becomes the Melvin magnetic field deformed $AdS_6 \times S^4$.  Although the solution represents the D-branes under the Melvin RR one-form we use a simple observation to see that it also describes the solution of D-branes under the Maxwell magnetic field. As the magnetic field we consider is the part of the background itself we have presented an alternative to previous literature, because our method does not require the assumption of negligible back reaction.  Next, we use the found solution to investigate the meson property through D4/D8 system (Sakai-Sugimoto model) and compare it with those studied by other authors.   Finally, we present a detailed analysis about the Wilson loop therein and results show that the external Maxwell magnetic field will enhance the quark-antiquark potential.

\vspace{2cm}
\begin{flushleft}
*E-mail:  whhwung@mail.ncku.edu.tw\\
\end{flushleft}
\newpage
\section{Introduction}
The holographic gauge/gravity correspondence has been used extensively to investigate properties of strongly coupled gauge theories [1-4].  This method has also been studied in various external conditions, including nonzero temperature [3] and background electric and magnetic fields [5-10], in which it exhibits many properties that are expected of QCD.

The background magnetic field are particularly interesting in that they may be physically relevant in neutron stars.  The background magnetic fields have also some interesting effects on the QCD ground state.  The basic mechanism for this is that in a strong magnetic field all the quarks sit in the lowest Landau
level, and the dynamics are effectively 1+1 dimensional, where the catalysis of chiral symmetry breaking was demonstrated explicitly, including the Sakai-Sugimoto model [8-11].

A method for introducing a background magnetic field has been previously discussed in the D3/D7 model in ref. [5] (The approach was first used to study drag force in SYM plasma [12].).  The author [5] consider pure gauge B field in the supergravity background, which is equivalent to exciting a gauge field on the world-volume of the flavor branes.    This is because that the general DBI action is
$$ S_{DBI} =  \int d^{p+1}8\xi~ e^{-\phi}\sqrt {-Det(G_{ab} + B_{ab} + 2\pi \alpha' F_{ab})}$$
$$ + \int \sum_r C_r\wedge exp(B_{ab} + 2\pi \alpha' F_{ab}). \eqno{(1.1)}$$
Here $G_{ab}$ is the induced metric and $B_{ab}$  the B-field on the Dp probe brane world volume.  $F_{ab}$ is its world volume gauge field and $C_r$ is the RR field. 

A simple way to introduce magnetic field would be to consider a pure gauge B-field along parts of the D3-branes  world volume, e.g.:  $B^{(2)} = H dx^2 \wedge dx 3$.   Since $B_{ab}$ can be mixed with the gauge field strength $F_{ab}$ this is equivalent to a magnetic field on the world volume.  Also, as the B-field is pure gauge, dB = 0, the corresponding background is still a solution to the supergravity equation of motion. 

   Despite that the above observation is really very simple it does suffer some intrinsic problems in itself.
\\

$\bullet $  First, although adding a pure gauge B-field does not change the backgrounds of the supergravity we does not know whether adding a pure Maxwell  F field strength will modify the background of metric, dilaton field or RR fields.

$\bullet $ Second, in considering  the F1 string moving on the background geometry (such as in investigating the Wilson loop property [13]) then, as described in the action (1.2),  the B-field is the gauge field to which a string can couple, the effect of B-field on the F1 string will be different from that coming from the Maxwell F field strength of potential $A_\mu$, in contrast to the D-string action described in (1.1). 
$$S_{Nambu-Goto} = \int d\tau d\sigma \sqrt {-Det(G_{ab}})+ \epsilon^{ab}\partial_a X^\mu\partial_b X^\mu B_{\mu\nu}  +  \int d\tau \dot X^\mu A_\mu. \eqno{(1.2)}$$

$\bullet $  Thus, it is useful to find an exact supergravity background which duals to holographic gauge theory with external Maxwell magnetic field.
\\

   To begin with, let us first make a comment about our previous finding about the Melvin magnetic field deformed $AdS_5 \times S^5$ [14].  In that paper we apply the transformation of mixing azimuthal and internal coordinate [15] to the 11D M-theory with a stack $N$ M2 branes [16,17] and then use T duality [18] to find the spacetime of a stack of  N D3-branes with external magnetic field.  In the near-horizon limit the background becomes the magnetic deformed $AdS_5 \times  S^5$ as followings.  
$$ds_{10}^2 =  \sqrt{1+B^2U^{-2}cos\gamma^2}\left[U^2(- dt^2+ dx_1^2+dx_2^2 +{1\over \sqrt{1+B^2U^{-2}}}dx_3^2)+{1\over U^2}dU^2 +\right.\hspace{2cm} $$
$$\left.\hspace{1cm} \left(d\gamma^2 + {cos^2\gamma\ d\varphi_1^2\over 1+ B^2U^{-2}cos^2\gamma} + sin^2\gamma\ d\Omega_3^2 \right)\right], ~~~~~with~~ A_{\phi_1} ={BU^{-2}~cos^2\gamma\over 2\left(1+ B^2U^{-2} cos^2\gamma\right)}.\eqno{(1.3)}$$ 
As the $A_{\phi_1}$ depend on the coordinate $U$ and not on the D3 brane worldvolume coordinates $t, x_1,x_2,x_3$ it is not a suitable background to describe those with external magnetic field.

In section II,  we first apply the transformation of mixing azimuthal with wrapped coordinate to the 11D M-theory with a stack N M5-branes [19] to find the spacetime of a stack of  N D4-branes with magnetic field in 10D IIA string theory, after the Kaluza-Klein reduction [20,21].  In the near-horizon limit the background becomes the inhomogeneously Melvin  magnetic field deformed  $AdS_6 \times S^4$.  

  In section III, we show that although the supergravity solution represents the D-branes under the external Melvin  RR one-form one can use a simple observation to see it also is the solution of D-branes under the external Maxwell electromagnetic field.  We argue that the EM field considered by previous authors [8-10] affects only the flavor sector and the color degrees of freedom do not sense this field.  The magnetic we consider  in this paper, however, are the part of the background itself.  Therefore, we have presented an interesting alternative to previous procedures because our method does not require the assumption of negligible back reaction.

  In section IV, we use the found supergravity background to investigate the holographic gauge theory with external Maxwell magnetic field flux.  We discuss the relations between using above supergravity solution and those using in recent by many authors [5-11] to investigate the holographic gauge theory with external magnetic field through D4/D8.  In section V we investigate the Wilson loop under the external Maxwell magnetic field.    The results show that magnetic field will enhance the quark-antiquark potential and may produce a negative potential energy.  

 The last section is devoted to a short conclusion. 
\section{Supergravity Solution with Melvin Magnetic Field}
The bosonic sector action of  D=11 supergravity is [19]
$$I_{11} = \int d^{11}\sqrt{-g}\left[R(g) -{1\over48}F^2_{(4)}+ {1\over6}F_
{(4)}\wedge F_{(4)} \wedge A_{(3)}\right].\eqno{(2.1)}$$
Using above action the full $N$ M5-branes solution  is given by [17]
$$ds^2_{11}=H^{-1\over3}\left(-dt^2+dz^2+dw^2+dr^2+ r^2d\phi^2+dx_5^2\right)+H^{2\over3}\sum_{a=1}^5 ~dx_a^2,$$
$$ A_{tzwr\phi x_5}= r(H^{-1}-1),\hspace{8cm}\eqno{(2.2)}$$
in which $H$ is the harmonic function defined by 
$$ H = 1+ {1\over R^{3}}, ~~~~~~~R^2\equiv \sum_{a=1}^5\left(x_a\right)^2.\eqno{(2.3)}$$

Following the Melvin twist prescription [15] we transform the angle $\phi$ by mixing it with the wrapped coordinate $x_5$ in the following substituting
$$\phi \rightarrow \phi + B x_5.\eqno{(2.4)}$$ 
Using the above substitution the line element (2.2)  becomes
$$ds_{11}^2 = H^{-1\over3} (1+B^2r^2)\left(dx_5^2+ {Br^2\over 1+B^2r^2}d\phi\right)^2 + H^{-1\over3}\left(-dt^2+dz^2+dw^2+dr^2+ {r^2\over 1+B^2r^2}d\phi^2\right)$$
$$ \hspace{8cm}+H^{2\over3}\sum_{a=1}^5dx_a^2. \eqno{(2.5)}$$
As the relation between the 11D M-theory metric and string frame metric, dilaton field and magnetic potential is described by 
$$ds_{11}^2= e^{-2\Phi/3}ds_{10}^2+  e^{4\Phi/3} (dx_5+2 A_\mu dx^\mu )^2 , \eqno{(2.6)} $$
the 10D IIA background is described by
$$ds_{10}^2 =\sqrt{1+ B^2 r^2} \left[H^{-1\over2}\left(-dt^2+dz^2+dw^2+dr^2+ {r^2\over 1+B^2r^2}d\phi^2\right)+H^{1\over2} \sum_{a=1}^5 dx_a^2 \right],\eqno{(2.7)}$$ 
$$e^\Phi = H^{-1/4}(1+B^2r^2)^{3/4},~~~~A_{\phi} ={Br^2\over 2\left(1+B^2 r^2\right)},~~~A_{tzwr\phi}= r(H^{-1}-1),\eqno{(2.8)}$$
in which $A_{\phi}$ is the Melvin magnetic potential and $A_{tzwr\phi}$ is the RR field.  In the case of  $B=0$  the above spacetime becomes the well-known geometry of a stack of  D4-branes. 

In the Horizon limit the above background becomes
$$ds_{10}^2 =\sqrt{1+ B^2 r^2} \left[U^{3\over2}\left(-dt^2+dz^2+dw^2+dr^2+ {r^2\over 1+B^2r^2}d\phi^2\right)+U^{-3\over2} \left(dU^2+U^2d\Omega_4^2\right) \right].\eqno{(2.9)}$$ 
The EM field tensor calculated from $A_{\phi}$ is
$$ F_{r\phi}= \partial_r A_{\phi}-\partial_\phi A_r = {Br\over \left(1+B^2 r^2\right)^2} ~~~\Rightarrow B_z(r) = F_{xy}= {B\over \left(1+B^2 r^2\right)^2}.\eqno{(2.10)}$$
In the case of  $B=0$  above spacetime becomes the well-known geometry of $AdS_6\times S^4$. Thus, this background describes the magnetic Melvin field deformed $AdS_6\times S^4$.  
 Let us make following comments to discuss above solution.

1. We see that there is a factor $(1+ B^2 r^2)$ in (2.9).  Thus the physical quantities evaluated in the dual gravity side by above supergravity solution will be {\bf not homogeneous on the x-y plane} ($x = r\cos\phi, ~ y= r\sin\phi$).  This reflects the inhomogeneity of the Melvin magnetic field in our solution.  

2.  To our knowledge,  the supergravity solution duals to holographic gauge theory with constant magnetic field has not yet been found.  We thus study the problem under inhomogeneous magnetic field and hope that the found properties could, more or less, also show in the system under a constant magnetic field. 

\section{Melvin RR Field vs. Maxwell Magnetic Field}
\subsection{Melvin RR Field}
Note that after the Kaluza-Klein reduction by (2.6) the  D=11 action (2.1) is reduced to the type IIA bosonic action which in the string frame becomes [20,21]
$$I_{IIA} = \int d^{10}\sqrt{-g}\left[e^{-2\Phi}\left(R(g)+4\nabla_M\Phi 
\nabla^M\Phi-{1\over12}F_{MNP}^{(NS)}F^{(NS)MNP}\right)\right.$$
$$\left. -{1\over48}F_{MNPQ}^{(RR)}F^{(RR)MNPQ}-{1\over4}{\cal F}^{(KK)}_{MN}{\cal F}^{(KK)MN}\right],\eqno{(3.1)}$$ 
in which ${\cal F}^{(KK)}_{MN}$ is the field strength of the Kaluza-Klein vector $A_\phi$ in (2.8).  Using above action we make following remarks :
\\

  1.  Above relation shows a distinguishing feature of the NS sector as opposed to the RR sector: the dilaton coupling is a uniform $e^{-2\Phi}$ in the NS sector, and it does not couple (in string frame) to the RR sector field strengths.  It is this property that we shall interprete the field strength ${\cal F}_{MN}^{(KK)}$ as the RR field strength and associated Kaluza-Klein vector as the RR one-form. Therefore our solution (2.7) shall be interpreted as the supergravity solution of  D4 branes  under external RR one-form which has a special function form in (2.8). 

     2. Also, as the NS-NS B field shown in the action is through the strength tensor $F^{(NS)}_{MNP}$, which becomes zero if B field is a constant value, we can introduce arbitrary constant B field without break the solution. In short, the supergravity solution is not modified by a constant B-field since $F^{(NS)}_{MNP} = dB = 0$ and does not act as a source for the other supergravity fields.  This property has been used in [22] to construct the supergravity solution duals to the non-commutative N = 4 SYM in four dimensions.

\subsection{Maxwell Magnetic Field}
Let us now introduce an external Maxwell electromagnetic field into the D4 branes system.  In this case the total action becomes 
$$I_{IIA} + I_{external~Maxwell~Field}= \int d^{10}\sqrt{-g}\left[e^{-2\Phi}\left(R(g)+4\nabla_M\Phi 
\nabla^M\Phi-{1\over12}F_{MNP}^{(NS)}F^{(NS)MNP}\right)\right.$$
$$\left. -{1\over48}F_{MNPQ}^{(RR)}F^{(RR)MNPQ}-{1\over4}{\cal F}^{(KK)}_{MN}{\cal F}^{(KK)MN}-{1\over4} {\cal F}^{(Maxwell)}_{MN}{\cal F}^{(Maxwell)MN}\right].\eqno{(3.2)}$$ 
We may use above action to find the supergravity solution under external Maxwell field.   However, as the Maxwell field strength (${\cal F}^{(Maxwell)}_{MN}$) has a same form as the RR Melvin field strength (${\cal F}^{(KK)}_{MN}$) the supergravity solution of  D4 branes under external RR Melvin field  strength (in the case of zero ${\cal F}^{(Maxwell)}_{MN}$) also describes the solution of the D4 branes under external  Maxwell field strength (in the case of zero ${\cal F}^{(KK)}_{MN}$).    It is this simple observation that {\bf although the supergravity solution represent the D4-brane under the external Melvin field flux of RR one-form we see that it also is the solution of D4-brane under the external Maxwell electromagnetic field while turn off the ${\cal F}^{(KK)}_{MN}$}.   In this paper we will use the found supergravity solution to investigate the dual gauge theory under the external Maxwell field.

   Note that a Melvin gauge field may be physically different from flavor gauge fields in the Sakai-Sugimoto model. The latter are generally non-Abelian, and external magnetic fields may be included for the charge generator embedded in the non-Abelian flavor group, thus coupling differently to different flavors, while the Melvin gauge field lives in the bulk and couples uniformly.  Therefore, the magnetic field considered by previous authors [5-11] affects only the flavor sector and the color degrees of freedom do not sense this field.  Thus,  the external fields are always viewed as some appropriate gauge mode on the probe itself and do not backreact or modify the background. The magnetic field we consider  in this paper, however, are the part of the background itself.   Thus, we have presented an interesting alternative to previous procedures because our method does not require the assumption of negligible back reaction.
\section{Meson in Sakai-Sugimoto Model under Maxwell Magnetic Field}
In this section we will use our gravity background to study the Sakai-Sugimoto model under the Maxwell magnetic field and compare it with that studied by previous author [8-11] 
\subsection{Meson under Melvin Maxwell Magnetic Field}
To investigate the Sakai-Sugimoto model [11] on the above supergravity background we consider a D8-brane embedded in the D4 branes background (2.9) with $U = U(w)$. Then the induced metric on the D8-brane is given by 
$$ds_{D8}^2 =\sqrt{1+ B^2 r^2} \left[\left(U^{3\over2} +U^{-3\over2} U'^2\right)dw^2+U^{3\over2}\left(-dt^2+dz^2+dr^2+ {r^2\over 1+B^2r^2}d\phi^2\right)+U^{1/2}d\Omega_4^2 \right].\eqno{(4.1)}$$ 
Using above  induced metric, $F_{r\phi}$ in (2.10) and dilaton field in (2.8) the DBI action (1.1) and associated Lagrangian for the probe D8 brane are
 $$S_{DBI}= \int d^{9}\xi~ e^{-\phi}\sqrt {-det(G_{ab} + F_{ab})} =  \int d^{9}\xi~{\cal L}(U,U'),\eqno{(4.2)}$$
in which 
$${\cal L} = (1+ B^2 r^2)~U^{5\over2}\sqrt{1+ {1\over U^3}\left({\partial U\over \partial w}\right)^2} \sqrt{U^3+ B_z(r)^2 },~~~~~~ B_z(r) \equiv {B\over \left(1+B^2 r^2\right)^2}.\eqno{(4.3)}$$
As ${\cal L}(U,U')$ is independent of $w$ the Hamiltonian corresponding to $w$ will be a constant.  Therefore
$$ {\cal H}_w = U' {\partial {\cal L}(U,U')\over \partial U'}- {\cal L}(U,U')= const.,\eqno{(4.4)}$$
and we find an equation
$$U^4 {\sqrt{1+B_z(r)^2/U^3}\over \sqrt{1+U'^2/U^3}}= U_0^4 \sqrt{1+B_z(r)^2/U_0^3},~~~~~~ B_z(r) \equiv {B\over \left(1+B^2 r^2\right)^2}\eqno{(4.5)}$$
in which $U_0$ is the minimum value of $U$ that the probe brane can reach satisfying $U'|_{U=U_0}=0$.   After numerical study  (4.5) we plot the function $U(w)$ in figure 1.  
\\
\\
\scalebox{1}{\hspace{5cm}\includegraphics{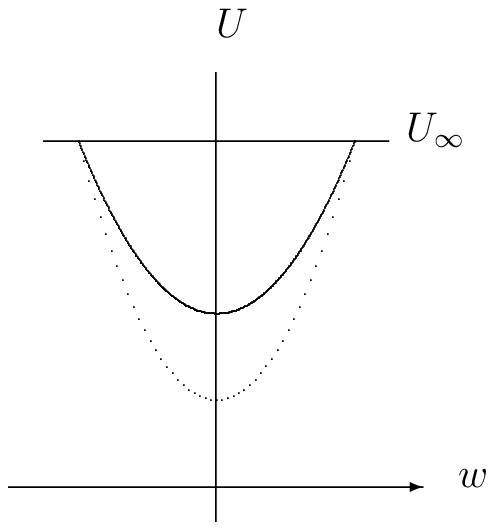}}
\\
\\
{\hspace{0cm} {\it Figure 1. The dotted U-shaped curve represents a profile in vanishing background field and the solid U-shaped curve represents a profile when a Melvin magnetic field.}
\\
\\
We see that the magnetic field will bend the profile of D8 branes and therefore force the brane pair to join closer to the boundary.    Therefore an external magnetic field will promote the chiral symmetry breaking, as that in [8].
\subsection{Meson under NS-NS B  Field}
As mentioned in section, as the  $B_{ab}$ can be mixed with the gauge field strength $F_{ab}$ in DBI action this is equivalent to a magnetic field on the world volume.  In this way the gravity background does not be modified by B (or Maxwell field) and the induced metric on the D8-brane used by many other authors [8-10] for the constant external magnetic field $B_0$ is given by 
$$ds_{D8}^2 =\left(U^{3\over2} +U^{-3\over2} U'^2\right)dw^2+U^{3\over2}\left(-dt^2+dz^2+dr^2+ r^2d\phi^2\right)+U^{1/2}d\Omega_4^2,\eqno{(4.6)}$$ 
which is different from our equation (4.1).  

The corresponding Lagrangian density can be easily calculated from DBI action, as before. As the Lagrangian density is independent of $w$ the Hamiltonian corresponding to $w$ will be a constant.  This then implies the equation
$$U^4 {\sqrt{1+B_0/U^3}\over \sqrt{1+U'^2/U^3}}= U_0^4 \sqrt{(1+B_0)^2/U_0^3},\eqno{(4.7)}$$
which becomes the equation (4.5) with simple substituting
 $$B_0 \rightarrow B_z(r).\eqno{(4.8)}$$  
It is surprised that although our metric (4.1) is very different from (4.6) the equation (4.5) and (4.8) have a very similar form.   Therefore the two method show a similar property that an external magnetic field will promote the chiral symmetry breaking, as that in [8].

Note that we could also follow the method of [11] to study the spectrum of a meson whose position is on a fixed position $r$.   In this case we need to consider the Wess-Zumino term in (1.1).     It is easy to see that simply substituting (4.8) to the result in [11] we could obtain the desired result in our model spacetime. 
\section{Wilson Loop }
In this section we will use our gravity background to study the Wilson loop under maxwell magnetic field.  Note that, as the previous study used the gravity background which does not be modified by B (or Maxwell field) they could not find the property presented in below.
\subsection{Wilson Loop : Case 1}
Following the Maldacena's computational technique the Wilson loop of a quark anti-quark pair is calculated from a dual string [13].  The string lies along a geodesic with endpoints on the $AdS_5$ boundary representing the quark and anti-quark positions.  

The first case we consider is that the quark and anti-quark  sit on $z=\pm{L\over2}$ with fixed $r$, as shown in figure 2. 
\\
\\
\scalebox{1}{\hspace{5cm}\includegraphics{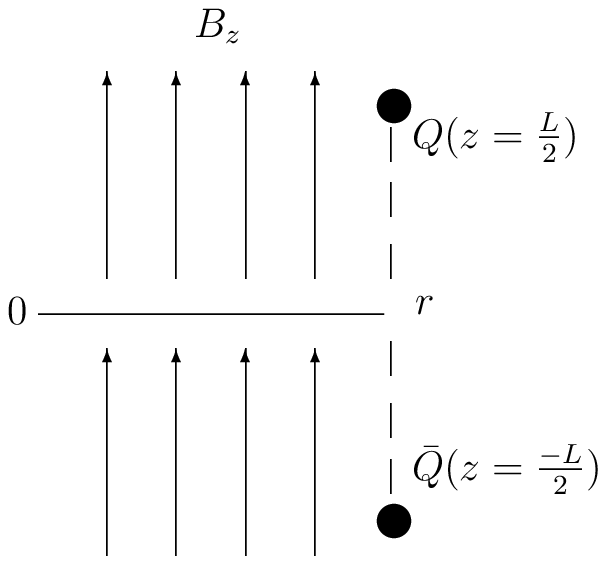}}
\\

{\it \hspace{2cm} Figure 2. The quark and anti-quark sit on $z=\pm{L\over2}$ with fixed $r$.}
\\

 Therefore, the string under the magnetic background (2.9) has following ansatz 
$$ t=\tau,~~~U=\sigma,~~~~z=z(\sigma),\eqno{(4.1)}$$
with a fixed value of $r$.   The Nambu-Goto action (1.2) becomes
$$S= {1\over 2\pi}\sqrt{1+B^2r^2} \int d\sigma \sqrt{1 +U^3(\partial_\sigma z)^2}.\eqno{(4.2)}$$
As the quark pair sits on the constant value of $r$ the  overall factor $(1+ B^2 r^2)$ is a just a constant the property of  the quark potential could be analyzed as before [13].  The method is reviewed  in below, as we need the formula in next subsection. 

  First, as the associated Lagrangian $\cal L$ does not depend on $z$ the momentum $\pi_z$ is a constant 
$$\pi_z \equiv {\partial {\cal L}\over \partial (\partial_\sigma z)}= \sqrt{1+B^2r^2} {U^3(\partial_\sigma z)\over \sqrt{1 +U^3(\partial_\sigma z)^2}}=\sqrt{1+B^2r^2}~U_0^{3/2},\eqno{(4.3)}$$
as at $U_0$ we have property $\partial_\sigma z \rightarrow\infty$.   Above relation implies that
$$(\partial_\sigma z)^2= {{1\over U^3}\over{U^3\over U_0^3}-1}.\eqno{(4.4)}$$
The distance $L$ between quark and antiquark is
$$L=2\int_0^{L/2} dz = 2\int_{U_0}^\infty d\sigma {dz\over d\sigma }= 2\int_{U_0}^\infty dU {{1\over U^{3/2}}\over \sqrt{{U^3\over U_0^3}-1}}= {2\over \sqrt {U_0}}\int_1^\infty dy {1\over y^{3/2}\sqrt{y^3-1}}= {1\over \sqrt {U_0}}{(\pi)^{3/2}\Gamma(5/3)\over \Gamma(5/6)}.\eqno{(4.5)}$$
Next, using (4.10) the interquark potential $V(U_0)$ could be calculated as follow 
$$V(U_0) ={ \sqrt{1+B^2r^2}\over \pi}\left[\int_{U_0}^\infty dU \sqrt{1+U^3(\partial_\sigma z)^2}- \int_0^\infty dU~ \right]\hspace{6cm} $$
$$= {\sqrt{1+B^2r^2}\over \pi}\left[\int_{U_0}^\infty dU \sqrt{1 +{1\over {U^3\over U_0^3}-1}}- \int_0^\infty dU\right]= -  {\sqrt{\pi}~U_0\Gamma({2\over3})\over \Gamma({1\over6})}\sqrt{1+B^2r^2},\hspace{0.5cm}\eqno{(4.6)}$$
in which we have subtracted the bare string energy.  Eqs (4.5) and (4.6) implies
$$V(L)= - \sqrt{1+B^2r^2}~{\Gamma({2\over3})\over \Gamma({1\over6})}{(\pi)^{7/2}\Gamma(5/3)^2\over \Gamma(5/6)^2}{1\over L^2},\eqno{(4.7)}$$
and we see that the Maxwell magnetic field could enhance  the quark-antiquark potential.
\subsection{Wilson Loop : Case 2}
The second case we consider is that  the quark and anti-quark sit on $x=\pm{L\over2}$, as shown in figure 3.  
\\
\\
\scalebox{1}{\hspace{5cm}\includegraphics{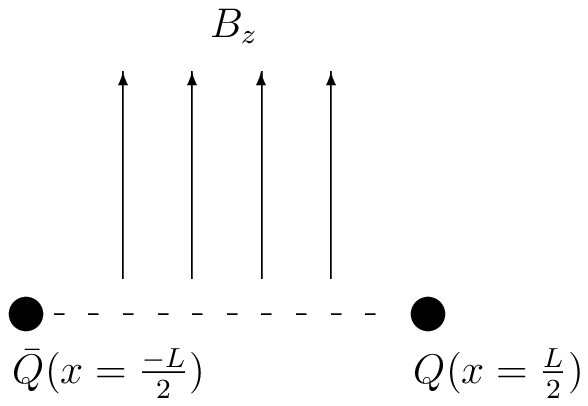}}
\\
\\

{\hspace{2cm} {\it Figure 3. The quark and anti-quark sit on $x=\pm{L\over2}$}
\\

Therefore, the string under the magnetic background (2.9) has the following ansatz 
$$ t=\tau,~~~U=\sigma,~~~r = r(\sigma).\eqno{(4.8)}$$
The Nambu-Goto action (1.2) becomes
$$S= {1\over 2\pi} \int d\sigma \sqrt{(1+B^2r^2)(1 +U^3(\partial_\sigma r)^ 2)},\eqno{(4.9)}$$
As we can not exactly solve this case we will consider the case with small $B$ field.  The action to first order of $B^2$ becomes
$$S \approx {1\over 2\pi} \int d\sigma \left[\sqrt{1 +U^3(\partial_\sigma r)^ 2}+{B^2r^2\over 2}\sqrt{1 +U^3(\partial_\sigma r)^ 2}\right].\eqno{(4.10)}$$
The second term is the corrected energy which could be calculated from the following formula
$$ \delta V  = {1\over \pi}\left[\int_{U_0}^\infty dU \sqrt{1 +U^3(\partial_\sigma r)^2}~{B^2r^2\over 2}- \int_0^\infty dU~~{B^2(L/2)^2\over 2}\right].\eqno{(4.11)}$$
Notice that, as the space is inhomogeneous we have to subtract the energy  of bare string  which is located at $r=L/2$.

  The functions  $(\partial_\sigma r)^2$ and $r$ in the first term of (4.11) are the zero-order functions and we can use  (4.4) and (4.5) to find their values (with $z\rightarrow r$ and $B\rightarrow  0$)  
$$(\partial_\sigma r)^2= {{1\over U^3}\over{U^3\over U_0^3}-1},~~r = \int_0^r dr = \int_{U_0}^U d\sigma {dr\over d\sigma }= \int_{U_0}^U dU {{1\over U^{3/2}}\over \sqrt{{U^3\over U_0^3}-1}}= {1\over \sqrt{U_0}}\int_1^{U/U_0} dx {1\over x^{3/2}\sqrt{x^3-1}}.\eqno{(4.12)}$$
After the substitutions (4.12) into (4.11) we find that 
$$ \delta V = {B^2\over 2\pi}\left[\int_{U_0}^\infty dU\left({{U^{3/2}\over U_0^{3/2}}\over \sqrt{ {U^3\over U_0^3}-1}}\left({1\over U_0}\int_1^{U\over U_0} dx {1\over x^2\sqrt{x^3-1}}\right)^2-(L/2)^2\right)- \int_0^{U_0} dU~(L/2)^2\right]  $$
$$ = {B^2\over 2\pi}\left[\int_1^\infty dy\left({y^{3/2}\over \sqrt{ y^3-1}}\left(\int_1^y dx {1\over x^{3/2}\sqrt{x^3-1}}\right)^2-\left({(\pi)^{3/2}\Gamma(5/3)\over \Gamma(5/6)}\right)^2\right)- \left({(\pi)^{3/2}\Gamma(5/3)\over \Gamma(5/6)}\right)^2\right] $$
$$=- 0.3785~B^2 ,\hspace{8cm}\eqno{(4.13)}$$
after numeric evaluations.  Thus, the Maxwell magnetic field will produce a negative potential energy, which also enhance  the quark-antiquark potential. 
\section{Conclusion}
In this paper we have constructed the supergravity background of  inhomogeneously magnetic field deformed $AdS_6 \times S^4$.   We have used a simple observation to see that the supergravity solution, which represents the D-brane under the external Melvin field flux of RR one-form, is also the solution of D-brane under the external Maxwell field flux.  We  use the solution to study the meson property through D4/D8 system and compared it with those studied by many authors [5-11].  As the magnetic field we consider is the part of the background itself we have presented an interesting alternative to previous procedures, because our method does not require the assumption of negligible back reaction.   We also present a detailed analysis about the Wilson loop therein.  The results show that the effect of magnetic field is to enhance the quark-antiquark potential and its effect will depend on the direction of field with respect to the direction of  quark-antiquark. 
\\
\\
{\bf Acknowledgments} :We are supported in part by the Taiwan National Science Council under grants 98-2112-M-006-008-MY3. 
\\
\\
\\
\\
\begin{center} {\bf  \Large References}\end{center}
\begin{enumerate}
\item J.~M. Maldacena, ``The large N limit of superconformal field theories  and supergravity,''  Adv. Theor. Math. Phys.  2  (1998) 231-252  [hep-th/9711200].
\item E.~Witten, ``Anti-de Sitter space and holography,'' Adv.\ Theor.\ Math.\ Phys.\   2 (1998) 253 [hep-th/9802150].
\item E.~Witten, ``Anti-de Sitter space, thermal phase transition, and confinement in  gauge theories,'' Adv.\ Theor.\ Math.\ Phys.\   2 (1998) 505 [hep-th/9803131]. 
\item S.~S.~Gubser, I.~R.~Klebanov and A.~M.~Polyakov, ``Gauge theory correlators from non-critical string theory,'' Phys.\ Lett.\ B 428 (1998) 105
[hep-th/9802109].
\item V. G. Filev, C. V. Johnson, R. C. Rashkov and K. S. Viswanathan, ``Flavoured large N gauge theory in an external magnetic field,"  JHEP 0710 (2007) 019 [hep-th/0701001], V. G. Filev,``Criticality, Scaling and Chiral Symmetry Breaking in External Magnetic Field," JHEP 0804 (2008) 088,  arXiv:0706.3811 [hep-th].
\item T. Albash, V. G. Filev, C. V. Johnson and A. Kundu, ``Finite Temperature Large N Gauge Theory with Quarks in an External Magnetic Field,"  JHEP 0807 (2008) 080,arXiv:0709.1547 [hep-th]; J. Erdmenger, R. Meyer and J. P. Shock, ``AdS/CFT with Flavour in Electric and Magnetic Kalb-Ramond Fields, " JHEP 0712, 091 (2007), arXiv:0709.1551 [hep-th].
\item  S. Penati, M. Pirrone and C. Ratti, ``Mesons in marginally deformed AdS/CFT," JHEP 0804 (2008) 037, arXiv:0710.4292 [hep-th]; K. D. Jensen, A. Karch, J. Price,``Strongly bound mesons at finite temperature and in magnetic fields from AdS/CFT,"  JHEP0804 (2008) 058, arXiv:0801.2401 [hep-th].
\item  C. V. Johnson and A. Kundu, ``External Fields and Chiral Symmetry Breaking in the Sakai-Sugimoto Model ,"   JHEP 0812 (2008)053, arXiv:0803.0038 [hep-th];  E. G. Thompson and D. T. Son,``Magnetized baryonic matter in holographic QCD,"  Phys.Rev.D78 (2008) 066007, arXiv:0806.0367 [hep-th].   
\item O. Bergman, G. Lifschytz and M. Lippert,`` Magnetic properties of dense holographic QCD," arXiv:0806.0366 [hep-th]; A. Rebhan, A. Schmitt and S. A. Stricker,`` Meson supercurrents and the Meissner effect in the Sakai-Sugimoto model," arXiv:0811.3533 [hep-th].
\item A. Karch and E. Katz, `` Adding flavor to AdS/CFT," JHEP 0206 (2002) 043 [hep-th/0205236];   M. Kruczenski, D. Mateos, R. C. Myers and D. J. Winters,``Meson Spectroscopy in AdS/CFT with Flavour," JHEP 0307 (2003) 049 [hep-th/0304032].
\item T. Sakai and S. Sugimoto, ``Low energy hadron physics in holographic QCD," Prog. Theor. Phys. 113 (2005) 843-882 [ hep-th/0412141]; ``More on a holographic dual of QCD,"  Prog. Theor. Phys. 114 (2006) 1083 [ hep-th/0507073]; C. V. Johnson and A. Kundu," Meson Spectra and Magnetic Fields in the Sakai-Sugimoto Model", arXiv:0904.4320 [hep-th].
\item T. Matsuo, D. Tomino and W. Y. Wen,``Drag force in SYM plasma with B field from AdS/CFT," JHEP0610 (2006) 055 [hep-th/0607178].
\item  J.~M. Maldacena,  ``Wilson loops in large N field theories,''  Phys.   Rev. Lett.  80 (1998) 4859-4862 [hep-th/9803002]; S.-J. Rey and J.-T. Yee,  ``Macroscopic strings as heavy quarks in large  N gauge theory and anti-de Sitter  supergravity,''   Eur. Phys. J.   C22 (2001) 379--394 [hep-th/9803001]
\item Wung-Hong Huang, `` Semiclassical Strings in Electric and Magnetic Fields Deformed $AdS_5 \times S^5$ Spacetimes," Phys.Rev.D73 (2006) 026007 [hep-th/0512117 ]; ``Spin Chain with Magnetic Field and Spinning String in Magnetic Field Background," Phys.Rev. D74 (2006) 027901 hep-th/0605242; ``Giant Magnons under NS-NS and Melvin Fields," JHEP0612 (2006) 040 [hep-th/0607161].  
\item M.A. Melvin, ``Pure magnetic and electric geons,'' Phys. Lett. 8 (1964) 65; 
F.~Dowker, J.~P.~Gauntlett, D.~A.~Kastor and J.~Traschen, ``The decay of magnetic fields in Kaluza-Klein theory,'' Phys.\ Rev.\ D52 (1995) 6929 [hep-th/9507143]; M.~S.~Costa and M.~Gutperle, ``The Kaluza-Klein Melvin solution in M-theory,'' JHEP 0103 (2001) 027 [hep-th/0012072].
\item C.~G.~Callan, J.~A.~Harvey and A.~Strominger, ``Supersymmetric string solitons,'' [hep-th/9112030].
\item  A.~Dabholkar, G.~W.~Gibbons, J.~A.~Harvey and F.~Ruiz Ruiz, ``Superstrings And Solitons,'' Nucl.\ Phys.\ B  340 (1990) 33; G.T. Horowitz and A.~Strominger, ``Black strings and P-branes,'' Nucl.\ Phys.\ B  360 (1991) 197.
\item P. Ginsparg and C. Vafa, Nucl. Phys. B289 (1987) 414; T. Buscher, Phys. Lett. B159 (1985) 127; B194 (1987) 59; B201 (1988) 466; S. F. Hassan, ``T-Duality, Space-time Spinors and R-R Fields in Curved Backgrounds,'' Nucl.Phys. B568 (2000) 145 [hep-th/9907152].
\item E. Cremmer, B. Julia and J. Scherk, Phys. Lett. B 76, 409 (1978).
\item I.C. Campbell and P.C. West, Nucl. Phys. B 243, 112 (1984); F. Giani and M. Pernici, Phys. Rev. D 30, 325 (1984).
\item  K.S. Stelle, ``BPS Branes in Supergravity," [hep-th/9803116].
\item  A. Hashimoto and N. Itzhaki, ``Non-Commutative Yang-Mills and the AdS/CFT Correspondence," Phys.Lett. B465 (1999) 142 [hep-th/9803116];  J. M. Maldacena and G. Russo, ``Large N Limit of Non-Commutative Gauge Theories ," JHEP 9909 (1999) 025 [hep-th/9908134].
\end{enumerate}
\end{document}